\documentclass[12pt]{article}
\usepackage{amsmath,amssymb,mathrsfs, multirow,multicol,bm}
\usepackage{graphics}
\usepackage{epic}
\usepackage{epsfig}
\usepackage{subfigure}
\usepackage{chngcntr}
\usepackage{color}

\newtheorem{thm}{Theorem}[section]

\newtheorem{rema}[thm]{Remark}

 {\ \\ {\bf Proof #1. }}%
 {\hfill\mbox{\rule{2 true mm}{3 true mm}}}

\numberwithin{equation}{section}
\def\theequation{\thesection.\arabic{equation}}
\def\be{\begin{eqnarray}}
\def\ee{\end{eqnarray}}
\def\numero{\refstepcounter{equation} (\theequation)}

\let\text=\textstyle

\def\be{\begin{eqnarray}}
\def\ee{\end{eqnarray}}
\def\ben{\begin{eqnarray*}}
\def\een{\end{eqnarray*}}
\def\bei{\begin{itemize}}
\def\eei{\end{itemize}}

\title{\bf Log-moment estimators for the    generalized  Linnik and  Mittag-Leffler distributions   \\with applications to financial modeling}

\author{Dexter O. Cahoy \\ Department of Mathematics and Statistics\\
       University of Houston-Downtown\\
                    Houston, TX  77002\\
         \texttt{cahoyd@uhd.edu} \\
\\
  Wojbor A. Woyczy\'nski \\ Department of Mathematics, Applied Mathematics \\
and Statistics, and Center for Stochastic  \\ and
Chaotic Processes in Science and Technology\\
     Case Western Reserve University\\
                    Cleveland, OH  44106\\
         \texttt{waw@case.edu}}

\date{August 3, 2018}

\begin{document}

\maketitle



\newpage

  {\bf Abstract:}     We propose formal  estimation procedures  for the parameters of the  generalized, heavy-tailed three-parameter Linnik $gL(\alpha,\mu, \delta)$  and  Mittag-Leffler $gML(\alpha,\mu, \delta)$ distributions.  The paper also aims to provide guidance about the different  inference procedures for the different  two-parameter Linnik and Mittag-Leffler  distributions in the current literature. The estimators are derived from the moments of the log-transformed random variables, and are shown to be asymptotically unbiased.  The estimation algorithms  are computationally efficient and the proposed procedures are   tested   using    the daily S\&P 500  and Dow Jones  index data. The results show that the two-parameter Linnik and  Mittag-Leffler models are not flexible   enough to accurately model the current stock market data.   



\section{Introduction}
 
In recent years, the   heavy-tailed  two-parameter Linnik  L$(\alpha, \lambda)$ distribution (see, e.g., Kotz and Ostrovskii ( 1996) )  introduced in  Linnik (1963), defined by the characteristic function
$$
\phi(t) = \left( 1 +  |\lambda t|^\alpha  \right)^{-1},
$$ 
where  $\lambda>0$ is the scale parameter, $t \in \mathbb{R}$, and $0 < \alpha \leq 2$, 
has gained popularity in many applications. For instance, it has been used to model  discrete-time stationary processes  particularly in finance (e.g., S\&P 500 index, see Kozubowski (1999, 2001)). In addition, extensive theoretical studies of the distribution has been carried out in Devroye,
(1990), Kozubowski (2001), Kotz and Ostrovskii (1996),  Lin (1998), Pakes (1998),  Cahoy (2012), Gunaratnam and Woyczynski (2015), Gorska  and Woyczynski (2015) and in the references cited therein.       Recall that the  L$(\alpha, \lambda)$  distribution is a   geometric stable distribution(Klebanov et al., 1985; Halvarsson, 2013), that is, it is invariant under random summation with the random number of summands determined by the geometric distribution. 

The parameter estimation problem for $\alpha$, when $\lambda=1$, was addressed by Anderson (1992) using the methods of Leitch and Paulson (1975),  Paulson et. al (1975), and 
Press (1972). Jacques et. al (1999) adopted Press (1972)'s technique to estimate the parameters $\alpha$ and $\lambda$.  Similarly, Kozubowski (2001) suggested the fractional moment estimators while Cahoy (2012) derived  closed-form expressions of the point and interval estimators of the parameters   $\alpha$ and $\lambda.$  Note also that Cahoy, Uchaikin and Woyczynski  (2010) developed inference procedures for the  two-parameter Mittag-Leffler distribution with Laplace transform $\phi(t) = \left( 1 +  (\lambda t)^\alpha  \right)^{-1}.$

The main goal of this paper is to estimate the  parameters  of the heavy-tailed three-parameter generalized Linnik    family of one-dimensional distributions,  $gL(\alpha,\delta, \mu),$    with the characteristic  function  
\begin{equation}
\phi(t) = \left( \frac{\mu}{\mu+\vert t \vert^\alpha} \right)^\delta,  \qquad -\infty<t<\infty, \label{cfgl}
\end{equation}
 with $\mu>0, \delta >0$, and $0<\alpha\le 2$.  Another objective of this paper is to   estimate parameters  of the heavy-tailed three-parameter  generalized Mittag-Leffler distribution   $gML(\alpha,\delta, \mu)$ (see, e.g., Laskin (2003)) with the Laplace transform   
 $$
 \phi(t) = \left( \frac{\mu}{\mu+  t^\alpha} \right)^\delta, \qquad t \geq 0,\quad  0< \alpha \leq 1,
 $$ 
 with the corresponding density function
\begin{equation}
f(x) \;  =   \;  \mu^\delta x^{\delta \alpha -1} E_{\alpha, \delta \alpha}^\delta (-\mu x^\alpha), \qquad x > 0,
\end{equation}
where
\begin{equation}
	E_{\beta,\gamma}^\eta(z) = \sum_{r=0}^\infty \frac{(\eta)_r \; z^r}{r!\Gamma(\beta r+\gamma) }, 
			\quad \beta,\gamma,\eta, z \in \mathbb{C}, \: \Re(\beta)>0,  
\end{equation} 
is the generalized Mittag--Leffler function (see, e.g., Cahoy and Polito (2013)), with $(\eta)_r = \eta( \eta+1)\ldots ( \eta + r-1), \eta \ne 0$,  representing the classical Pochhammer symbol.  We emphasize that estimation procedure  for   $gML(\alpha,\delta, \mu=1)$  was developed in Cahoy (2013).

Note that if  $\alpha=1$  and the data support is  $\mathbb{R}^+$ , we obtain  the gamma distribution.     When $\alpha=\delta=1$ and and data support is  $\mathbb{R}^+$,  then we obtain  the exponential distribution. It can be shown that  $gML(\alpha,  \delta, \mu)$ is a mixture of  generalized gamma densities with the strictly $\alpha^+$-stable density as the mixing distribution.    With $\delta=1$,  we have the usual  Mittag-Leffler distribution (see. e.g., Pillai (1990)) which  can be interpreted  as a  mixture of Weibull densities.

Finally, we also   compare the efficiency of the above three-parameter  models  with the existing models  (see e.g., Kozubowski (1999, 2001)) using stock market S\&P 500 and Dow Jones index data.

The paper is organized  as follows: In Section 2, we  provide structural representations of the generalized Linnik $gL(\alpha, \delta,  \mu)$  and the  generalized Mittag-Leffler $gML(\alpha, \delta,  \mu)$ random variables. In Section  3,  we derive the method-of-moments estimators based on the log-transformed data.  In Section 4, we test the algorithms using synthetic data. Section 5 shows the analyses of the $S\&P$ 500 and Dow Jones data.    We conclude in Section 6 with a discussion of  the key points of this work and possible future extensions of our study.


\section{Mixture representations and moments}

In this section we provide representations for random variables with generalized  Mittag-Leffler and Linnik distributions employing the standard L\'evy $\alpha$-stable random variables and review related results for completeness.

\subsection{Generalized Mittag-Leffler distributions on the positive half-line}

  {\bf Theorem 1.}
{\it 
  Let $0< \alpha < 1$, $S$ be a strictly $\alpha^+$-stable  random variable  with the Laplace transform $\exp ( - t^\alpha), t>0$, $U$  be an independent random variable with gamma   distribution   (with rate $\mu>0$,  and shape parameter $\delta>0$) , i.e., with density function
			\begin{equation}
				f_U(u)=\frac{\mu^\delta}{\Gamma (\delta)} u^{\delta-1} e^{-\mu u}, \quad u>0.
			\end{equation}
   Then the random variable 
\begin{equation}
X \; =  \; U^{ 1 /\alpha } S    \label{gmlx}
\end{equation}
 has the $gML(\alpha,\delta, \mu) $  distribution.}
 
 \medskip
 The proof  is straightforward:  
 
 \[
\phi_X(t) = \mathbf{E}  e^{ i t (    U^{1 / \alpha} S_\alpha )}  =   \mathbf{E}  e^{  -t^\alpha     U }   
       = \left( 1 + \frac{ t^\alpha}{\mu}  \right)^{-\delta}.   
\]

 \medskip
 
The proof can also be found, for example, in Pakes (1998). Recall that the $\alpha^+$-stable random variable  can be conveniently generated using the classical Kanter (1975) formula 
\begin{equation}
S \stackrel{d}{=}  \frac{\sin(\alpha U)[ \sin((1-\alpha)U)]^{1/\alpha-1}}{[\sin (U)]^{1/\alpha}E^{1/\alpha-1}}, 
\end{equation}
where $U$  is uniformly distributed in $[0,\pi ]$,   and $E$ is a  standard exponential random variable (with rate/scale parameter one)  independent of $U.$   The $q$-th fractional moment of $X$ can be  easily derived from the above result, and is given below.

\bigskip

\noindent  {\bf Remark 1. } {\emph{As $\alpha \to 1$ or $S \equiv 1$,   the $gML(\alpha,\delta, \mu) $  converges (in distribution) to the gamma distribution with rate parameter $\mu$ and shape parameter $\delta.$} 

\bigskip

\noindent {\bf Theorem 2. }
{\it 
Let $0< \alpha < 1.$  Then
\[ 
\mathbf{E} X^q \; = \;  \frac{\Gamma (\delta + q/ \alpha) \Gamma (1- q/\alpha) }{\mu^{q / \alpha} \; \Gamma ( 1-  q )  \Gamma (\delta )},  \qquad 0<q < \alpha.  
\]
}

\medskip

The proof follows directly from the standard moment formulas 
\begin{equation}
\mathbf{E} U^{q/\alpha} = \frac{\Gamma (\delta + q/ \alpha) }{\mu^{q / \alpha}  \;  \Gamma (\delta )},  \qquad {\rm and}    \qquad  \mathbf{E} S^\alpha =\frac{ \Gamma (1- q/\alpha) }{ \Gamma ( 1-  q ) }.
\end{equation}
See also Cahoy and Polito (2013).

\bigskip

\noindent  {\bf Remark 2. } {\emph{As $\alpha \to 1$  or $S \equiv 1$,} 

\[ 
\mathbf{E} X^q \; = \;  \frac{\Gamma (\delta + q)  }{\mu^{q} \; \Gamma (\delta )},  \qquad 0<q < \alpha.  
\]

\subsection{Generalized Linnik distributions on the entire real line}

{\bf Theorem 3.}
{\it 
Let $0< \alpha \leq 2$, and $S_\alpha   $  be a random variable with a  symmetric  $\alpha$-stable distribution  with characteristic function  $\exp ( - |t|^\alpha)$,  and  $U$  be   an independent  gamma  distributed random variable  with  density (2.1).  
Then the  random variable 
\begin{equation}
Y \;= \;  U^{ 1 /\alpha } S_\alpha     \label{glx}
\end{equation}
has the $gL(\alpha,\delta, \mu)$
 distribution.
}

\medskip
 The proof  follows from the proof of Theorem 1.   Note that Devroye (1990) had the proof  for $\delta=1/\delta', \mu=1.$   Apparently, the case $\alpha =1$    is essentially different in both families.

 The symmetric $\alpha$-stable random variable $S_\alpha$ can be generated using the standard Chambers-Mallows-Stuck (1976) formula
 \begin{equation}
S_\alpha \stackrel{d}{=}   \frac{\sin(\alpha U_2)}{[ \cos(U_2) ]^{1/\alpha}} \left( \frac{\cos ( (1-\alpha)U_2 )}{E}\right)^{1/\alpha-1}, 
\end{equation}
where $U_2$  is  uniformly distributed on $[-\pi/2, \pi/2]$, and $E$ is  independent of $U_2$ and exponentially distributed with parameter one.     An expression for the $q$-th fractional moment  of $Y$ is  derived   below.


\noindent \newtheorem{cor1}{Proposition}
\begin{cor1}
Let $0< \alpha \leq 2$ and $Y \; \stackrel{d}{=}  \; gL(\alpha,\delta, \mu).$   Then
\[ 
\mathbf{E} |Y|^q \; =  \;   \frac{\Gamma(q) \; \sin (q) \; \Gamma ( \delta + q / \alpha) }{\mu^{q / \alpha} \; \sin (\pi q/ \alpha) \; \cos (\pi q/2) \; \Gamma ( \delta) \;  \Gamma (q/\alpha)}, \qquad 0<q < \alpha.  
\]
\end{cor1}
\noindent \textbf{Proof}. Note that 

\begin{equation}  
\mathbf{E} |Y|^q  \; = \; \mathbf{E}(U^{q / \alpha}) \;  \cdot \; \mathbf{E} |S_\alpha|^q .  
\end{equation}

Using the $q$-th fractional moment of the symmetric stable random variable $S_\alpha$ (see, Bening et al., 2004)
\begin{equation*}
  \mathbf{E} |S_\alpha|^q  = \frac{\Gamma (1-q/\alpha)}{\cos (q \pi/2) \Gamma (1-q)}   
\end{equation*}
we have
\begin{align} \notag
\mathbf{E} |Y|^q          &= \frac{\Gamma (\delta + q / \alpha)}{\mu^{q/\alpha} \Gamma ( \delta) }  \left( \frac{\Gamma (1-q/\alpha)}{\cos (q \pi/2) \Gamma (1-q)} \right). 
\end{align}
An application of  the reflection formula for the gamma function,
$\Gamma(1-p)\Gamma(p)=\pi/\sin(\pi p)$, completes the proof.


\section{Parameter estimation via the logarithmic moments}

\subsection{Generalized 3-parameter Mittag-Leffler distribution \\$gML(\alpha,\delta, \mu)$}

Following Cahoy, Uchaikin and Woyczynski  (2010), we apply the log transformation to the random variable $X$ given in (\ref{gmlx}) as 
\begin{equation}
X^{'} \stackrel{d}{=}  \frac{1}{\alpha}U^{'} + S^{'},   
\end{equation}
where $X^{'}= \ln (X)$, $U^{'}=\ln (U)$, and $S^{'}=\ln (S)$.   For reproducibility,  we can recall the first four log-moments of $S$ from Zolotarev (1986), and Cahoy, Uchaikin and Woyczynski (2010):
\[
\mathbf{E}\left(  S^{'} \right) = \mathbb{C}\left( \frac{1}{\alpha}-1\right), \qquad  \mathbf{E} \left( S^{'}\right) ^2=\left(\frac{1}{\alpha}-1\right)^2 \mathbb{C}^2 +
\frac{\pi^2}{6}\left( \frac{1}{\alpha^2}-1\right),
\]
\[
\mathbf{E} \left(S^{'}\right) ^3=\frac{-2(\alpha-1)^3\mathbb{C}^3+\mathbb{C}\pi^2
(\alpha-1)^2(1+\alpha)-4(\alpha^3-1)\zeta(3)}{2\alpha^3}, 
\]

\[
\mathbf{E} \left(S^{'}\right)^4 = \frac{1}{60}\bigg[
\bigg(\frac{1}{\alpha^3}-\frac{1}{\alpha^4}\bigg)\bigg(
60\mathbb{C}^4(\alpha-1)^3-60\mathbb{C}^2\pi^2(\alpha-1)^2(1+\alpha) \notag
\]
\[
\qquad +\pi^4(\alpha-3)(1+\alpha)(3+\alpha) +480\mathbb{C}(\alpha^3-1)\zeta (3)\bigg)
\bigg], 
\]
where $\mathbb{C}\simeq 0.5772$ is the Euler's constant.   

 It is straightforward to show the probability density of $U^{'}$ as
\begin{equation}
f(u^{'}) \; = \; \frac{\mu^\delta \exp \{\delta u' -\mu \exp (u')  \}}{\Gamma (\delta)}, \qquad u' \in \mathbb{R}.
\end{equation}
Using the   polygamma function of order  $k$, $ \psi^{(k)} (\delta) = \frac{d^{k+1}  \ln   \Gamma  (\delta) }{d \delta^{k+1}},$ the  first four log-moments of $U^{'}$ are 
\[
\mathbf{E}(U^{'})=\psi ( \delta ) - \ln \mu , \qquad \mathbf{E}
\left( U^{'} \right)^2= (\ln \mu- \psi ( \delta ) )^2 + \psi^{(1)} ( \delta ),
\] 
 \[
\mathbf{E}\left(  U^{'} \right)^3 = - \left( \ln \mu  -  \psi ( \delta )\right)^3 + 
 3 (\psi ( \delta ) - \ln \mu )\psi^{(1)} ( \delta ) + \psi^{(2)} ( \delta ),
\]
and
\[
\mathbf{E}\left( U^{'}\right)^4 = \left( \ln \mu \right)^4 -  4 (\ln \mu)(\psi ( \delta )) ^3 + (\psi ( \delta ) )^4 
\]
\[
+ 6 \left( \ln \mu \right)^2 \psi^{(1)} ( \delta ) + 3 ( \psi^{(1)} ( \delta ) )^2 
+ 6 ( \psi ( \delta ) ) ^2 \left(  (\ln \mu)^2  + \psi^{(1)} ( \delta ) \right)
\]
\[
 - 4 \psi ( \delta ) \left( (\ln \mu)^3 + 3  (\ln \mu ) \psi^{(1)} ( \delta )  - \psi^{(2)} ( \delta )  \right)  - 4  (\ln \mu ) \psi^{(2)} ( \delta ) +   \psi^{(3)} ( \delta ).
\]

Using the above moments,    the estimating equations  are as follows (see, also Cahoy and Polito (2013) where they were mentioned without showing the elementary (although tedious) algebra of moments):
			\[
				\mu_{X'} = \mathbf{E} \left( X' \right) =  \mathbb{C} \bigg( \frac{1}{\alpha} -1\bigg) +
				\frac{\psi ( \delta )-\ln (\mu)}{\alpha },
			\]
			\[
				\sigma_{X'}^2= \frac{\pi^2}{6}\bigg( \frac{1}{\alpha^2} -1\bigg) + \frac{1}{\alpha^2} \psi^{(1)} (\delta),
			\]
 and
			\[
				\mu_3= \mathbf{E} \left( X'- \mu_{X'}  \right)^3 =
				\frac{ \psi^{(2)} (\delta) - 2\left(\alpha^3 - 1\right)\zeta (3)}{\alpha^3},
			\]
where  $\zeta (\cdot)$ is the Riemann Zeta function.

Finally, using the estimators $\hat{\mu}_3$ and $  \hat{\sigma}_{X'}^2$, we can solve  the above equations for  the variance and the third central moment, perhaps using a numerical software to obtain the estimators  $\hat{\delta}$ and $\hat{\alpha}$.	Plugging $\hat{\alpha}$ and $\hat{\delta}$ into the mean equation above, we obtain the following estimator of the parameter $\mu$:
			\begin{equation}
				\hat{\mu}=\exp \left(- \left[\hat{\alpha} \left( \hat{\mu}_{X^{'}} -\mathbb{C} (1/\hat{\alpha}-1) \right)
				-\psi (\hat{\delta} ) \right] \right).
			 \end{equation}

\subsection{2-parameter  Mittag-Leffler distribution $gML(\alpha,1, \mu)$}
We start by emphasizing that this two-parameter version is different from what had been studied in Cahoy, Uchaikin and Woyczynski  (2010), which is  $gML(\alpha,1, \mu=\lambda^{-\alpha})$ in section 1,  and  from  Cahoy (2013), which is $gML(\alpha,\delta, \mu=1)$.  If $\delta =1$ then $\psi(1) = - \mathbb{C},  \psi^{(1)}(1) = \pi^2/6,$ and $\psi^{(2)}(1) = -2\zeta (3)$.  In addition, 
		$$
				\mu_{X'} = \mathbf{E} \left( X' \right) = 
				\frac{\ln (\mu)}{\alpha } - \mathbb{C} ,  \qquad 
				\sigma_{X'}^2= \frac{\pi^2(2-\alpha^2)}{6 \alpha^2} ,
		$$
		$$
				\mu_3= -2 \zeta (3), \qquad  {\rm and} \qquad  			
				\mu_4=\frac{\psi^{(4)}(1)  +2\pi^2\left(\alpha^2-2 \right) \zeta (3)}{\alpha^4}.
		$$

From the first two moments we obtain the following closed-form expressions of the  estimators of $\alpha$ and $\mu$:  
\begin{equation}
\hat {\alpha} = \frac{\sqrt{2} \pi}{\sqrt{6 \hat{\sigma}_{X'}^2 +\pi^2} }\qquad {\rm and}  \qquad  \hat{\mu} =  \exp \left(-\hat{\alpha} (\mathbb{C} + \hat{\mu}_{X^{'}} )  \right). 
\end{equation}
Note that these estimators are always non-negative as required and are asymptotically unbiased as shown in Proposition 2 below. 

\medskip

{\bf Proposition 2.} \emph{ Let $X_1, X_2, \ldots, X_n \; \;  \stackrel{iid}{=} \; gML(\alpha,1, \mu)$. Then}
\begin{equation}
\sqrt{n}\left(   \widehat{\alpha} - \alpha   \right) \stackrel{d}{\longrightarrow} \textsl{N}\left( 0\; ,\; \frac{ 36  \psi^{(4)}(1)  + 
 (\alpha^2 -2)  \left(72 \pi^2  \zeta (3) - \pi^4\right) }{36 \alpha^2 \mu^2}\right),  
\end{equation} 

\emph{and}

\begin{equation}
\sqrt{n}\left(  \widehat{\mu} - \mu  \right) \stackrel{d}{\longrightarrow} \textsl{N}\left( 0 \; ,\; \sigma_{\widehat{\mu} }^2\right),  \qquad  n \to \infty,
\end{equation} 
\emph{where }
 \begin{equation}
\sigma_{\widehat{\mu}}^2= 
\frac{ -6  \alpha^4 (\alpha^2-2) b^2 \pi^2 +    144 \alpha^5  \mu \ln (\mu) \zeta (3)  }{16 \mu^2 \pi^4}
\end{equation} 
$$
+\frac{  
  \ln(\mu)^2 (-(\alpha^2-2)^2 \pi^4 + 36  \psi^{(4)}(1)   + 
     72 (\alpha^2 -2) \pi^2 \zeta(3))}{16 \mu^2 \pi^4}.
$$

 {\bf Proof.}   The proof directly follows from the asymptotic normality of sample moments and the multivariate delta method,     where 
\begin{equation}
\sqrt{n}\big(\textbf{g}(\widehat{\bm{\theta}}_n)-\textbf{g}(\bm{\theta})\big)\stackrel{d}{\to} \textsl{N}\left( \bm{0},\; \bm{\dot{\textbf{g}}}(\bm{\theta})^{\text{T}}\bf{\Sigma}\bf{\dot{g}}(\bm{\theta})\right), 
\end{equation} 
with the variance-covariance matrix  
\begin{equation}
\bf{\Sigma} = \left(
       \begin{array}{cc}
         \sigma_{X^{'}}^2 & \mu_3  \\
         \mu_3  & \mu_4 -\sigma_{X^{'}}^4 \\
       \end{array}
     \right),
\end{equation} 
and 
\[
 \textbf{g}(\mu_{X^{'}},\sigma_{X^{'}}^2) \; = \; \left(  \; \frac{\sqrt{2} \pi}{\sqrt{6 \sigma_{X'}^2 +\pi^2} }  \quad, \quad    \exp \left(-\alpha (\mathbb{C} + \mu_{Y^{'}} )  \right) \right)^{\text{T}},
\]
 $\widehat{\bm{\theta}}_n=(\widehat{\mu}_{X^{'}},  \widehat{\sigma}_{X^{'}}^2)^{\text{T}},$  and   $\bm{\dot{\textbf{g}}}(\bm{\theta})= \nabla \textbf{g}(\bm{\theta})^{\text{T}}$ is the gradient matrix.   The above results can be used to approximate the $(1-\nu)100\%$ confidence intervals for  $\alpha$ and $\mu.$



\subsection{Generalized 3-parameter Linnik distribution $gL(\alpha,\delta, \mu)$}

Applying the log transformation to the absolute value of the generalized Linnik random variable $Y$ given in (\ref{glx}), we get an expression 
\begin{equation*}
Y^{'} \stackrel{d}{=}  \frac{1}{\alpha}U^{'} + S_\alpha^{'},   
\end{equation*}
where   $S_\alpha^{'}=\ln (|S_\alpha|)$.  The first four integer-order log-moments of $S_\alpha$  (see, Cahoy (2012)) are as follows: 
\[
\mathbf{E}\left(  S_\alpha^{'} \right) = \mathbb{C}\left( \frac{1}{\alpha}-1\right), \qquad  \mathbf{E} \left( S_\alpha^{'}\right)^2=
\frac{12 \mathbb{C}^2(\alpha-1)^2 + \left( \alpha^2 + 2 \right)  \pi^2 }{12 \alpha^2},  
\]
\[
\mathbf{E} \left(S_\alpha^{'}\right)^3= \frac{(1-\alpha) \left(4(\alpha-1)^2\mathbb{C}^3 + (\alpha^2+2)\mathbb{C}\pi^2 + 8 (\alpha^2 + \alpha +1) \zeta (3)\right)}{4 \alpha^3}
\]
and 
\[
\mathbf{E} \left(S_\alpha^{'}\right)^4 = \frac{1}{240 \alpha^4}\bigg[ 240 (\alpha-1)^4\mathbb{C}^4 + 120 (\alpha-1)^2( \alpha^2+2)\mathbb{C}^2 \pi^2  \notag
\]
\[
\qquad + \; (19\alpha^4 +20 \alpha^2 +36)\pi^4 + 1920 ( \alpha-1)^2(\alpha^2 + \alpha +1) \mathbb{C} \zeta (3) \bigg].
\]

The moments above  yield  the same mean $\mu_{Y'}$ and the centered third order moment $\mu_3$ as in the previous subsection. The  variance then can be calculated to be 
\begin{equation}
  \sigma_{Y^{'}}^2= \frac{\pi^2 (\alpha^2+2)}{12 \alpha^2}  + \frac{\psi^{(1)}(\delta)}{\alpha^2}. 
\end{equation}
Now the estimation approach employed  in the previous subsection for the generalized Mittag-Leffler distribution  $gML(\alpha,\delta, \mu)$  can also be applied in the present case.  The only difference here lies in the formula for the variance being used in the minimization process.

\subsection{2-parameter  Linnik distribution $gL(\alpha,1, \mu)$}
We start by emphasizing that this two-parameter version is different from what had been studied in  Cahoy (2012), which is $gL(\alpha,1, \mu=\lambda^{-\alpha})$ in the first un-numbered equation in section 1. If $\delta =1$ then 
		\[
				\mu_{Y'} = \mathbf{E} \left( Y' \right) = 
				\frac{\ln (\mu)}{\alpha } - \mathbb{C} , \; \; 
				\sigma_{Y'}^2= \frac{\pi^2(\alpha^2+4)}{12 \alpha^2} ,
			 \; 
				\mu_3= -2 \zeta (3), \;  \mu_4=\frac{A}{240 \alpha^4}, \; {\rm where} 	 
\]
\begin{eqnarray*}
				A &=&  1920 ( \alpha -1)^4  \mathbb{C}^4 + 
   480 (\alpha-1)^2 (2 + \alpha^2) \mathbb{C}^2 \pi^2 + (112 + 40 \alpha^2 + 
      19 \alpha^4) \pi^4 \\
& & +  \;  3840 (\alpha-1)^2 (1 + \alpha + \alpha^2)  \mathbb{C} \zeta(3). 
\end{eqnarray*}

Moreover, we obtain the following closed-form expressions of the  estimators of $\alpha$ and $\mu$:  
\begin{equation}
\hat {\alpha} = \frac{2\pi}{\sqrt{12 \hat{\sigma}_{Y'}^2 -\pi^2} }\qquad {\rm and}  \qquad  \hat{\mu} =  \exp \left(-\hat{\alpha} (\mathbb{C} + \hat{\mu}_{Y^{'}} )  \right). 
\end{equation}

{\bf Proposition 3.} \emph{ Let $Y_1, Y_2, \ldots, Y_n \; \;  \stackrel{iid}{=} \; gL(\alpha,1, \mu)$. Then}
\begin{equation}
\sqrt{n}\left(   \widehat{\alpha} - \alpha   \right) \stackrel{d}{\longrightarrow} \textsl{N}\left( 0\; ,\; \sigma_{\hat\alpha}^2 \right),  
\end{equation} 

\begin{equation}
 \sigma_{\hat\alpha}^2 = \frac{ 1440 ( \alpha-1)^4  \mathbb{C} ^4 + 
   360 (\alpha-1)^2 (2 + \alpha^2) \mathbb{C}^2 \pi^2 }{180 \alpha^2 \mu^2}
\end{equation} 
$$
+ \frac{  (64 + 20 \alpha^2 + 
      13 \alpha^4) \pi^4 +    2880 (\alpha-1)^2 (1 + \alpha + \alpha^2) \mathbb{C} \zeta(3)}{180 \alpha^2 \mu^2},
$$
\emph{and}

\begin{equation}
\sqrt{n}\left(  \widehat{\mu} - \mu  \right) \stackrel{d}{\longrightarrow} \textsl{N}\left( 0\; ,\; \sigma_{\widehat{\mu} }^2\right),  \qquad  n \to \infty,
\end{equation} 
\emph{where }
 \begin{equation}
\sigma_{\widehat{\mu}  }^2= 
\frac{15 \alpha^4 (4 + 
      \alpha^2) \mu^2 \pi^2 + (1440 (\alpha -1)^4 \mathbb{C}^4 + 
      360 (\alpha -1)^2 (2 + \alpha^2) \mathbb{C}^2 \pi^2 }{80 \mu^2 \pi^4}
\end{equation} 
$$
+\frac{  (64 + 20 \alpha^2 + 
         13 \alpha^4) \pi^4 \ln ( \mu)^2 + 
   720\log ( \mu) (\alpha^5 b +        4 (\alpha-1))^2 (1 + \alpha + \alpha^2) \mathbb{C}\ln ( \mu ) ) \zeta (3)}{80 \mu^2 \pi^4}.
$$
 
\medskip

 {\bf Proof.}   The proof directly follows from Proposition 2 above where
\[
 \textbf{g}(\mu_{Y^{'}},\sigma_{Y^{'}}^2) \; = \; \left(  \; \frac{2\pi}{\sqrt{12 \sigma_{Y'}^2 -\pi^2} } \quad, \quad   \exp \left(-\alpha (\mathbb{C} + \mu_{Y^{'}} )  \right) \right)^{\text{T}},
\]
and the components of the variance-covariance matrix  $\bf{\Sigma}$ are given in the beginning of this subsection. The above results can be used to approximate the $(1-\nu)100\%$ confidence intervals for  $\alpha$ and $\mu.$ 


\section{Testing our estimation procedures on simulated data}

In this section we will test the performance of our estimators  using simulated data.    Furthermore,  to quantify the performance errors we will calculate   the  mean bias, 
$$
MB=  {\rm Mean}( |\hat{\theta}-\theta|/\theta),
$$
  and the coefficient of variation 
   $$
  CV =   {\rm StandardDeviation}(\hat{\theta}) / {\rm Mean}( \hat{\theta})  
  $$
   for our estimators based on  1000 generated data samples for different parameter values.  

\subsection{Generalized Mittag-Leffler distribution $gML(\alpha,\delta, \mu)$}

For reproducibility, we used the \texttt{optim} function in R to minimize  $ ( \sigma_{X'}^2 -  \hat{\sigma}_{X'}^2)^2 +  ( \mu_3- \hat{\mu}_3)^2$ with respect to $\alpha $ and $\delta$  using the initial value $(0.1, 1)$. Note that expressions for $\sigma_{X'}^2$ and  $\mu_3$ are in Section 3.1.  We emphasize that other built-in functions in R  like the \texttt{polygamma} are used as well in  the calculation process. However, the stable random variables are generated following Kanter’s formula (2.3) and C-M-S formula (2.6) due to their elegance. 

 The point estimates of $\hat{\alpha}$ and $\hat{\delta}$ are then plugged in the point estimator   $\hat{\mu}.$    From Table 1,  the bias of $\hat \mu$ is around 19\% when $n=10^3$ and is around $6\%$ when $n=10^4.$ The CV fluctuates around 7.6\% when $n=10^4.$  Generally, Table 1 indicated positive results for the proposed method.

\bigskip

\noindent {\bf  Table 1: }{\it  The mean bias and CV of the proposed estimators  for the  $gML(\alpha,\delta, \mu)$ family  using three different values of $\alpha$,  $\delta=0.5$, and $\mu=1$,    for sample sizes $n=10^2, 10^3, 10^4$.}

 		\begin{table}[h!t!b!p!] 
		
	 			\centering
			\begin{tabular}{cc|ccc|ccc}
				\multicolumn{2}{c}{}  &  \multicolumn{3}{c}{Bias}  &  \multicolumn{3}{c}{CV} \\
				$\alpha$ & $Est$ & $n=10^2$ & $10^3$ & $10^4$ & $n=10^2$ & $10^3$ & $10^4$ \\ \cline{1-8}
				\multirow{3}{*}{$0.5$}
				&  $\hat{\alpha}$    & 0.177  & 0.067  & 0.021   &  0.286 & 0.105 & 0.027   \\
				&  $\hat{\delta}$ & 0.340  & 0.116  & 0.037   &  0.366 &  0.143& 0.047     \\
				&  $\hat{\mu}$ & 0.607 & 0.193  & 0.063   &  0.558 &  0.231 & 0.080     \\ \cline{1-8}
				\multirow{3}{*}{$0.7$}
				&  $\hat{\alpha}$    & 0.162 & 0.065   & 0.020   &  0.282 & 0.094 & 0.026   \\
				&  $\hat{\delta}$ & 0.323  & 0.113   & 0.037   &  0.378 &  0.140 & 0.045    \\
				&  $\hat{\mu}$ & 0.568  & 0.191  & 0.061  &  0.569 &  0.232 & 0.076     \\ \cline{1-8}
				\multirow{3}{*}{$0.95$}
		&                   $\hat{\alpha}$    & 0.143  & 0.059   & 0.018   &  0.260 & 0.090 & 0.024   \\
				&  $\hat{\delta}$ & 0.299  & 0.111   & 0.034   &  0.301 &  0.137 & 0.043     \\
				   &  $\hat{\mu}$ & 0.536 & 0.190   & 0.058   &   0.474 &  0.229 & 0.073     \\ \cline{1-8}
                                    \end{tabular}

		\end{table}

\subsection{Generalized Linnik distribution $gL(\alpha,\delta, \mu)$}

In this subsection we are providing results from testing our estimation procedures for the parameters in the   $ gL(\alpha,\delta, \mu)$ family. The approach is similar to the one we used for the generalized Mittag-Leffler distributions. The initial value pair   used is $(\alpha_0, \delta_0) = (1,1).$  We also  calculated the  same statistics for comparison.   From Table 2, the bias went down to as  little as 2.4\%  and went as high as 9.8\% when $n=10^4.$  The CV ranges from 3.2\% to 12.4\%. Note that  the results   for $n=100$ suggest    larger samples are needed or better optimization procedure (like the L-BFGS-B method in R).   Also, the estimator for $\mu$ seems to get worse as the true $\alpha$ value approaches  two.  Overall, Table 2  provided  favorable results for the proposed method especially for large samples. 
Note that in practice  one can use bootstrap to quantify the variability of these estimators.   

 \newpage

\noindent {\bf  Table 2: }{\it The mean bias and  CV of the proposed estimators for the $gL(\alpha,\delta, \mu)$ distribution using three different values of $\alpha$,  $\delta=0.5$, and $\mu=1$,    for sample sizes $n=10^2, 10^3, 10^4$.}

 		\begin{table}[h!t!b!p!]

			\centering
			\begin{tabular}{cc|ccc|ccc}
				\multicolumn{2}{c}{}  &  \multicolumn{3}{c}{Bias}  &  \multicolumn{3}{c}{CV} \\
				$\alpha$ & $Est$ & $n=10^2$ & $10^3$ & $10^4$ & $n=10^2$ & $10^3$ & $10^4$ \\ \cline{1-8}
				\multirow{3}{*}{$0.6$}
				&  $\hat{\alpha}$    & 0.209  & 0.085   & 0.024  &  0.385 & 0.154 & 0.046   \\
				&  $\hat{\delta}$ & 0.360  & 0.129   & 0.039   &  0.393 &  0.160 & 0.052     \\
				&  $\hat{\mu}$ & 0.630  & 0.213   & 0.066   &  0.596 &  0.250 & 0.084     \\ \cline{1-8}
				\multirow{3}{*}{$1.2$}
				&  $\hat{\alpha}$    & 0.278  & 0.077   & 0.025   &  0.575 & 0.144 & 0.032   \\
				&  $\hat{\delta}$ & 0.528  & 0.135   & 0.044   &  0.878 &  0.168 & 0.056     \\
				&  $\hat{\mu}$ & 0.936  & 0.232   & 0.076   &  1.240 &  0.281 & 0.095     \\ \cline{1-8}
				\multirow{3}{*}{$1.8$}
				&  $\hat{\alpha}$    & 0.230  & 0.098   & 0.031  &  0.222 & 0.150 & 0.040   \\
				&  $\hat{\delta}$ & 1.507  & 0.009   & 0.056   &1.377   &  0.220 & 0.071     \\
				&  $\hat{\mu}$ & 1.123  & 0.024   & 0.098   & 1.507  &  0.372 & 0.124     \\ \cline{1-8}
                                    \end{tabular}
 		\end{table}

\section{Generalized Mittag-Leffler and Linnik distributions in modeling of financial data}

We applied the proposed models to the stock market data obtained from {\tt finance. yahoo.com.}  The Yahoo file contained the following variables about the daily index:  {\tt date,  open,  high,    low,  close, adj.close,  volume,}   but  we restricted  our calculations  to the daily high and adjusted closing indices to illustrate the  proposed models. Of course,  similar procedures can be applied to the rest of the dataset. The S\&P 500 dataset  covers the period  from January 3, 1950 to  August 30, 2017,   while Dow Jones  contains the information from January 29, 1985 to  August 30, 2017.  Our  analysis was thus based on   $17, 025$ daily S\&P 500  data points, and  $8,215$ Dow Jones Industrial Average  indices.   In the entire analyses,   we generated 1000 bootstrap samples to calculate the point and the 95\% interval estimates  of the parameters.  We also used the boundary corrected kernel density estimate  of the {evmix} package of R to compare  the fits of  $gML(\alpha,\delta, \mu)$  and  $gML(\alpha,1, \mu)$ whenever possible.  R scripts   are   available from the authors  upon request. 
 
\subsection{Standard and Poor's (S\&P) 500 index} 
 
It was originally called the "Composite Index"   when it was first  introduced as a  stock market index in 1923.  Three years later, the Composite Index expanded to 90 stocks, and then in 1957 --  to its current 500, and renamed  S\&P 500 Index.   It was the first index to be published daily. It contains 500 of the largest stocks in the United States.  It is a   benchmark for gauging  the overall health of the large American companies,  and  the U.S. stock market in general.  More than \$7.8 trillion is benchmarked to the index (Source: \emph{Investopedia}).


\subsubsection{Comparison between $gML(\alpha,\delta, \mu)$ and $gML(\alpha,1, \mu)$ distributions}

We fitted the   $gML(\alpha,\delta, \mu)$ to  the absolute values of the negative adjusted closing log returns ($n=9005$) from the S\&P 500 data.  Table 3   clearly indicates  that     $\alpha$ is favored to be less than one and $\delta$ to be larger than unity, which suggests that  a two-parameter    Mittag-Leffler model  is not adequate for this data.  This observation is reinforced by the two-parameter estimates from the same table. In particular,   $\hat\alpha>1$ despite the relatively large sample size. The estimates of $\mu$  are however similar.

\bigskip

\noindent {\bf  Table 3. }{\it  Parameter estimates for  $gML(\alpha,\delta, \mu)$ and $gML(\alpha,1, \mu)$ models applied to  (S\&P) 500  data.}

\begin{table}[h!t!b!p!]
			\centering
			\begin{tabular}{c||cc|cc}
				 
				   $Estimator$ & Point  & $95\%$ CI &    Point ($\delta=1$) & $95\%$ CI ($\delta=1$)  \\ 
\hline
				  $\hat{\alpha}$ & 0.993  &  (0.983, \; 1.003)   & 1.047 &    (1.038, \; 1.056)\\
				  $\hat{\delta}$ & 1.163  & (1.117, \; 1.212)  &  & \\
				  $\hat{\mu}$ &180.017&  (170.500,  \;  188.255 )  & 183.470&  (176.104, \; 191.190) \\
\hline 

  \end{tabular}
 
				\end{table}

To examine the model fit, we simulated data (sample size $2n=18,010$)   from the estimated model above.  Specifically,  we superimposed  the boundary corrected  kernel density estimates of the simulated data  on the histogram   of the observed data.   Figure 1  below shows the  good fit of the proposed model to the  daily negative adjusted closing S\&P  500  log returns.  The graph  demonstrates the advantage of flexibility of  the three-parameter model as opposed  to the two-parameter   $gML(\alpha, 1, \mu)$ distribution in capturing the peak  near the origin. With $\hat\alpha>1$,  plotting the fit of $gML(\hat \alpha, 1, \hat \mu)$  is meaningless and computationally impossible.

\begin{figure}[h!t!b!p!]
\centering{
  \includegraphics[ height=2in, width=4in]{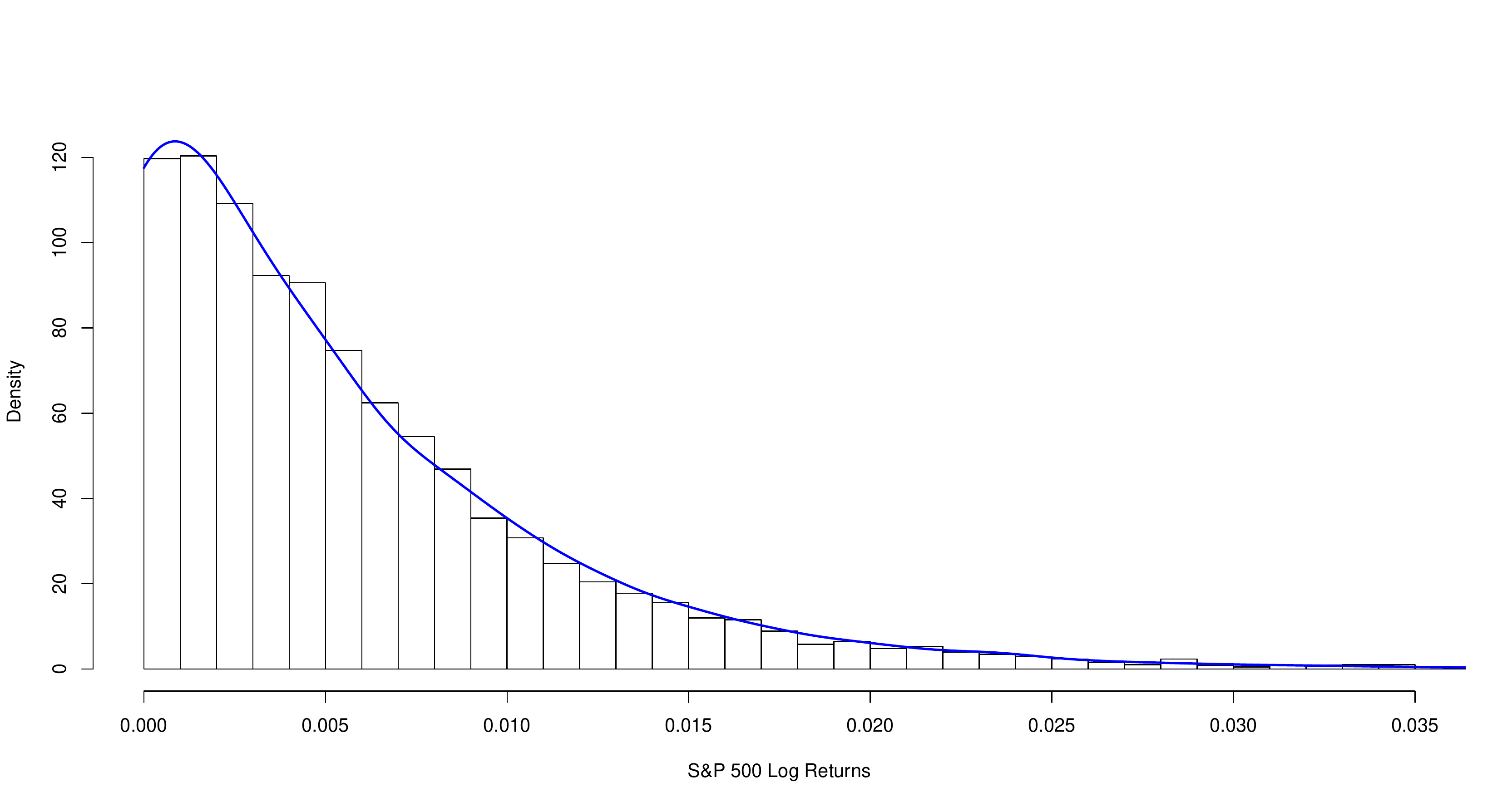}
        \caption{The histogram (using 150 bins) of the observed S\&P 500  data and the kernel density plot (bandwidth = 0.001) of the simulated data using the obtained estimates.}} \label{f1}
\end{figure}


\newpage 
\subsubsection{Comparison between $gL(\alpha,\delta, \mu)$ and $gL(\alpha,1, \mu)$ distributions}

We also analyzed the entire  log adjusted closing  returns ($n=17, 025$).  From the the estimates in Table 4,   the estimates for $\delta$ favor values larger than one, which implies that the daily  S\&P 500   log returns (adjusted closing) are not adequately described by  the two-parameter   Linnik distribution ($\delta=1$).  Note that we are not able to get an interval estimate for $\delta$ as the {\tt optim} function gives the same value as a root for every bootstrap sample.  The table also indicates that $\alpha$ is likely to be less than two and $\hat{\mu}$ is way larger than the estimate obtained in the preceding section.  The two-parameter estimate of $\alpha$ exceeds two indicating a bad fit of the model to the data. But the estimates of $\mu$ from both two- and three-parameter models are  comparable.

\bigskip

\noindent {\bf  Table 4. }{\it  Parameter estimates for  $gL(\alpha,\delta, \mu)$ and $gL(\alpha,1, \mu)$ models applied to  (S\&P) 500  data.}

 		\begin{table}[h!t!b!p!]
			\centering
\label{t4}
			\begin{tabular}{c||cc|cc}
				 
				   $Estimator$ & Point   & $95\%$ CI   &  Point ($\delta=1$) & $95\%$ CI ($\delta=1$) \\ 
\hline
				  $\hat{\alpha}$    & 1.915 & (1.881,\; 1.952)  &2.445 & (2.364, \; 2.529)    \\
				  $\hat{\delta}$ & 1.23  &     & & \\
				  $\hat{\mu}$ & 19115.36  &  (16255.82 ,\; 22675.40) &193059  & (131357.3 , \; 289028)  \\
\hline 
                                    \end{tabular}
 	 		 
		\end{table}

Figure 2  below confirms the good fit of the $gL(\alpha,\delta, \mu)$ family (using $2n= 33798$ simulated observations) to the  log adjusted closing  returns. It also reveals that the flexibility of the proposed three-parameter $gL(\alpha,\delta, \mu)$  permits better  capturing of  the peak  of  the data at the origin. Note that the algorithm used in the calculation was not able to generate a comparable fit as $\hat \alpha$ is way larger than two (upper bound).

\begin{figure}[h!t!b!p!]
\centering{
  \includegraphics[height=2in, width=4in]{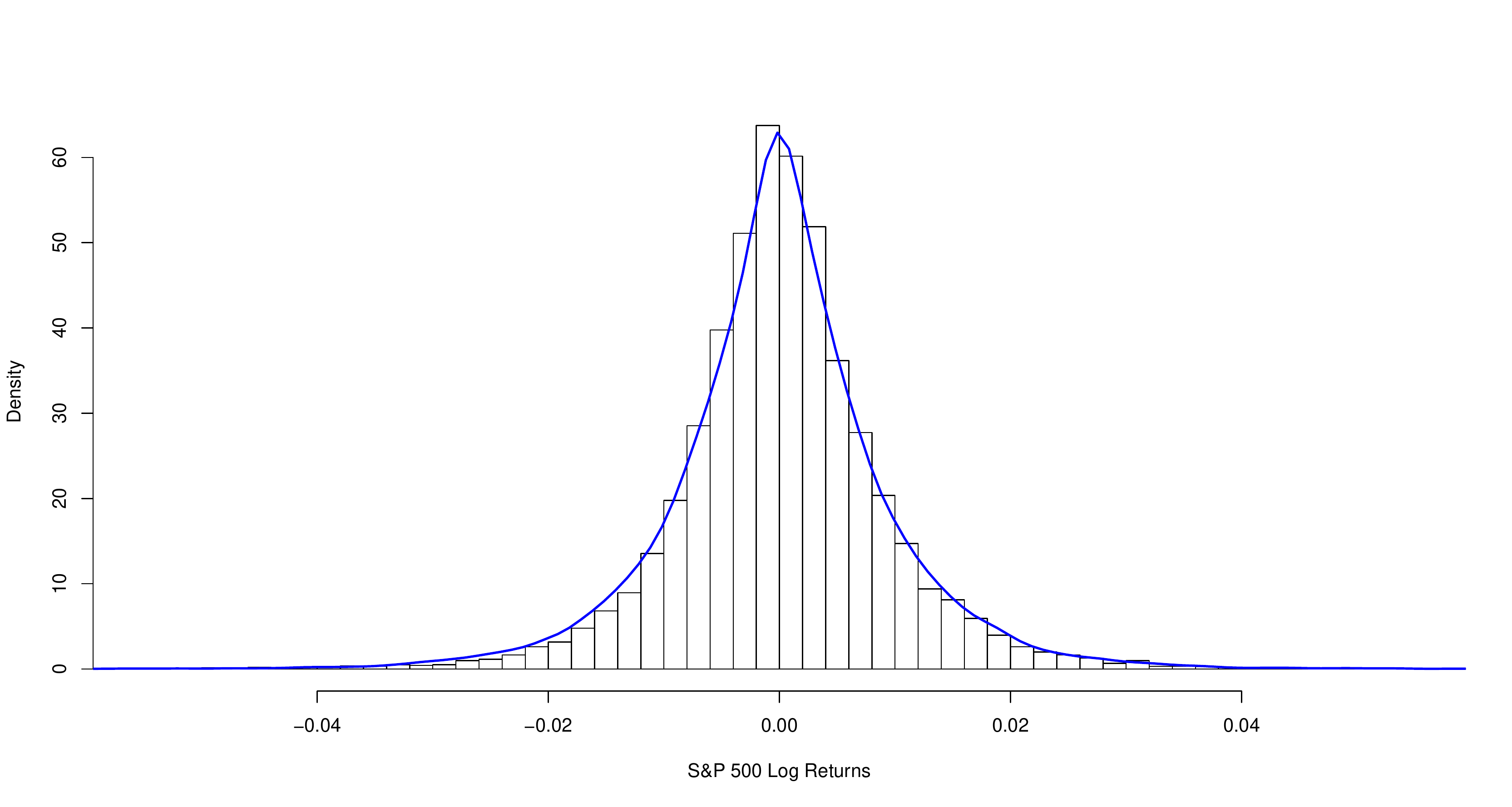} 
    \caption{\emph{ The histogram (using 150 bins) of the observed data and the kernel density plot (bandwidth = 0.001) of the simulated data using the obtained estimates superimposed on top.}} \label{f2}
}
\end{figure}



\subsection{Dow Jones Industrial Average index}

The Dow Jones Industrial Average (DJIA) is a price-weighted average of 30 significant stocks traded on the New York Stock Exchange  and the NASDAQ. The DJIA was invented by Charles Dow back in 1896.  Often referred to as "the Dow," the DJIA is one of the oldest, single most-watched indices in the world and includes companies such as General Electric Company, the Walt Disney Company, Exxon Mobil Corporation and Microsoft Corporation.      When the index was first launched, it included   companies that were almost purely industrial in nature. The first components  included railroads, cotton, gas, sugar, tobacco and oil companies. General Electric is the only one of the original Dow components that is still a part of the index in 2016. (Source: Investopedia)


\subsubsection{Comparison between $gML(\alpha,\delta, \mu)$ and  $gML(\alpha,1, \mu)$ distributions} 
 
The analysis here is similar to the one we carried out in the previous subsection and deals with  the  absolute values of the  negative adjusted closing  log returns($n= 4,359$) from the Dow Jones index. The 95\% CI for $\mu$ is between  154 and  173.   The  estimates of $\alpha$ strongly favor values less than  unity. The  point and interval estimates of $\delta$  indicate   that $\delta >1,$ which implies superiority of the generalized   $gML(\alpha,\delta, \mu)$ distribution for the  absolute values  of Dow daily Jones log returns over the   $gML(\alpha,1, \mu)$ distribution.  Observe that the   $gML(\alpha,1, \mu)$ fit provides similar estimates for $\mu$ but not  for $\alpha,$ and  that its kernel density estimate is missing as $\hat \alpha$ exceeds one.

 \bigskip

\noindent {\bf  Table 5. }{\it  Parameter estimates for  $gML(\alpha,\delta, \mu)$ model applied to  Dow Jones  data.}

		\begin{table}[h!t!b!p!]
			\centering
			\begin{tabular}{c||cc|cc}
				 
				   $Estimator$ & Point    & $95\%$ CI &  Point ($\delta=1$) & $95\%$ CI ($\delta=1$) \\ 
\hline
				  $\hat{\alpha}$    &0.983 & (0.974, \; 0.995 )  & 1.050 &  (1.036, \; 1.061)    \\
				  $\hat{\delta}$ & 1.211  &   (1.153,\; 1.274)  & & \\
				  $\hat{\mu}$ &  162.596  &  (153.578, \;  173.375) &165.952&  (157.062, \; 174.868)  \\
\hline 
                                    \end{tabular}
 			 		 
		\end{table}

Again,  as in the S\&P 500 case discussed above,  we  constructed the  graphs  (using $2n=8,718$ simulated observations) to investigate the model adequacy.  The smoothed density of the  $gML(0.983, 1.211, 162.596)$  is in Figure 3 .   It  basically confirms  what we have already observed above, that is,  the  three-parameter model  provides more flexibility  in capturing the peak of the 'cupping' near the origin than the two-parameter    Mittag-Leffler distributions.

\begin{figure}[h!t!b!p!]
\centering{
  \includegraphics[height=2in, width=4in]{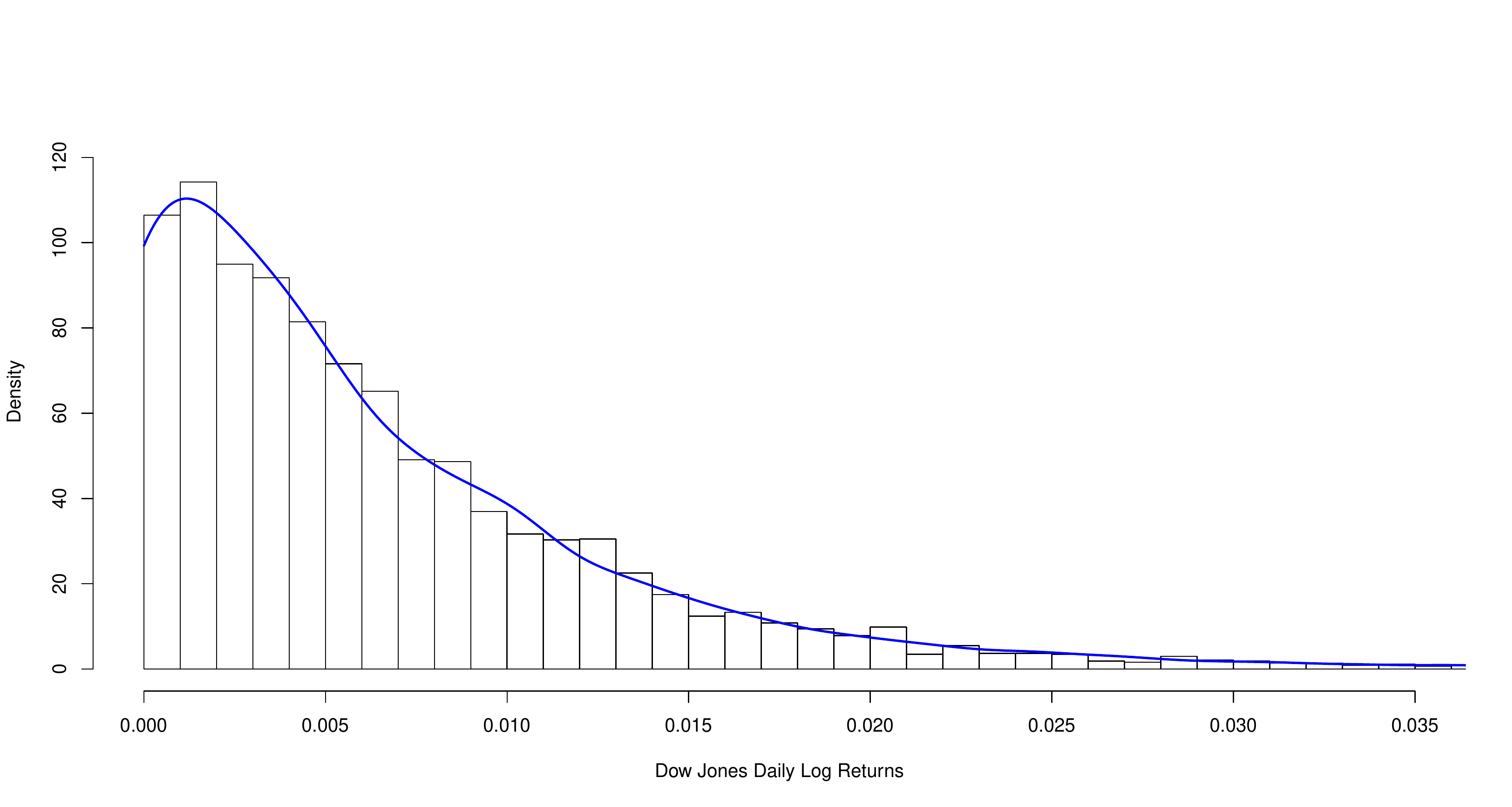}
 
   \caption{The histogram (using 150 bins) of the observed data and the kernel density plot (bandwidth = 0.001) of the simulated data using the obtained estimates superimposed on top.}} \label{f3}
\end{figure}

\subsubsection{Comparison between $gL(\alpha,\delta, \mu)$  and $gL(\alpha,1, \mu)$ distributions}

We also applied the generalized Linnik distribution to the  whole  adjusted closings   of the  daily Dow Jones log returns ($n= 17, 025$).   Looking at the estimate  of $\delta,$   $\hat \delta>1$, it is clear that   the  daily Dow Jones log returns (using the adjusted closing) cannot be adequately described by the      two-parameter Linnik distribution ($\delta=1$).    Moreover, the table below shows   $\alpha$ to be    likely less than 1.9.  


 \noindent {\bf  Table 6. }{\it  Parameter estimates for  $gL(\alpha,\delta, \mu)$ model applied to  Dow Jones  data.}

 		\begin{table}[h!t!b!p!]
			\centering
			\begin{tabular}{c||cc|cc}
				 
				   $Estimator$ & Point Estimate  & $95\%$ CI  &  Point ($\delta=1$) & $95\%$ CI ($\delta=1$) \\
\hline
				  $\hat{\alpha}$    &  1.844 & (1.803, 1.890)   & 2.258 & (2.158,  2.366)    \\
				  $\hat{\delta}$ & 1.24  &   &&  \\
				  $\hat{\mu}$ &11762   &  ( 9685  , 14605  ) &63024  & (39507 ,  103585 )     \\
\hline 
                                    \end{tabular}
 			 		 
		\end{table}

Figure 4  shows the fits of the   $gL(\alpha,\delta, \mu)$  and    $gL(\alpha,\delta=1, \mu)$ models.  Notice that the algorithm was able to produce a comparable fit in this case even if $\alpha >2.$   Furthermore, it validates   the  previous observation that the two-parameter    $gL(\alpha,\delta=1, \mu)$   is not adequate for the description of the daily Dow Jones adjusted closing log returns especially  in  capturing the peak  at the origin.

\begin{figure}[h!t!b!p!]
\centering{
   \includegraphics[height=3in, width=4in]{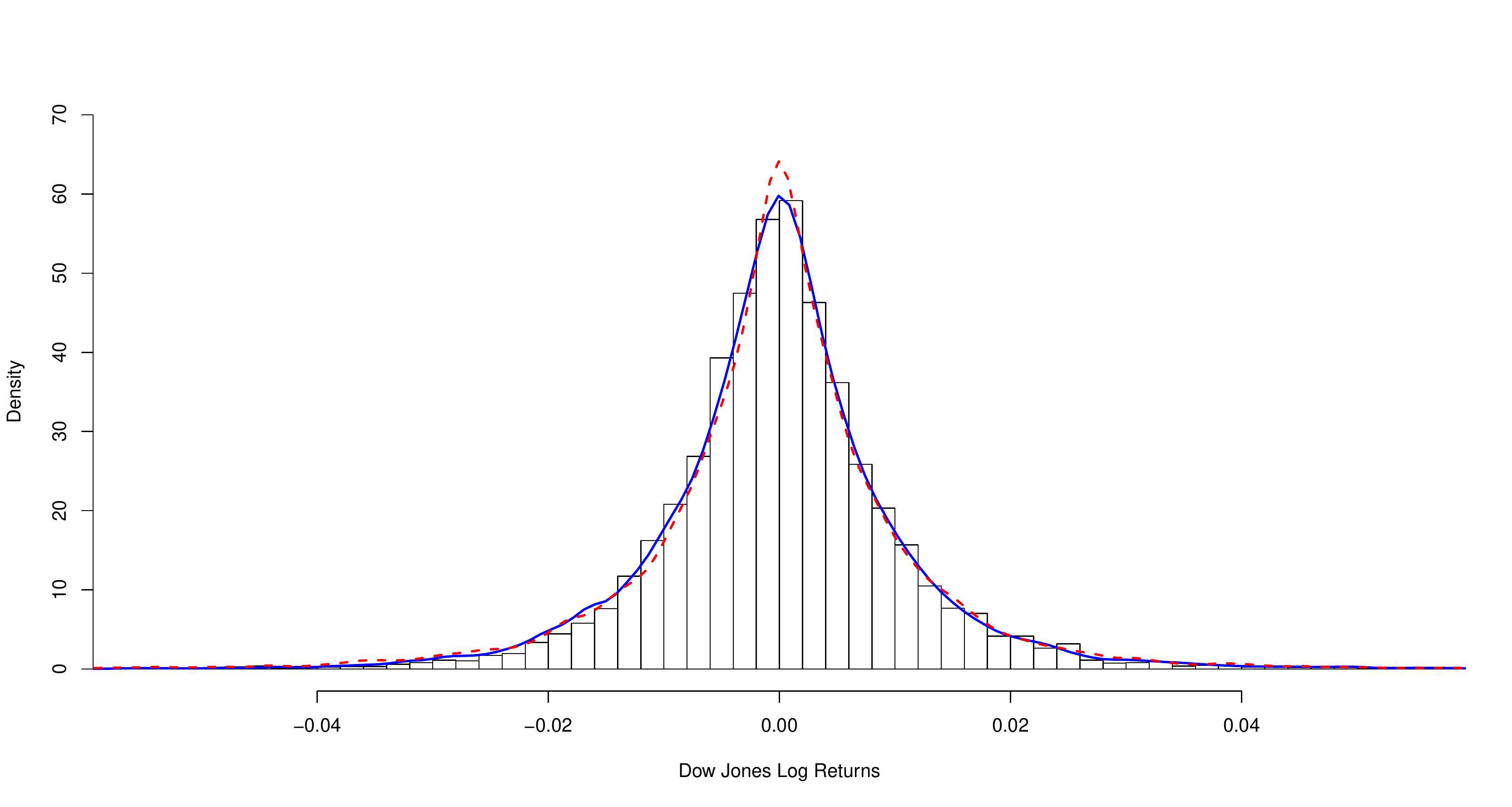}
  \caption{ The histogram (using 150 bins) of the observed data and the kernel density plots (bandwidth = 0.001) of the simulated data (red dashed for 2-parameter $gL(\alpha,\delta=1, \mu)$  and  solid blue for 3-parameter $gL(\alpha,\delta, \mu)$ ) using the obtained estimates.}}  
\end{figure}

\section{Concluding remarks}

The article proposes formal statistical inference  procedures for the  heavy-tailed three-parameter generalized Linnik  and  three-parameter generalized Mittag-Leffler families   of distributions.  The models provide considerable flexibility in modeling stationary discrete-time processes.    The consistency and unbiasedness of the point estimators were computationally tested and seemed to be acceptable.   Furthermore,   the structural representations and the  random number generation algorithms  were  provided for convenience.   
The paper provides  guidance to how to distinguish different subcases of these models that exist in the literature.

The heavy-tailed three-parameter generalized Linnik  and generalized Mittag-Leffler models present evidence that the adjusted S\&P 500  and Dow Jones  log returns can obey these probabilistic laws. The comparison of the proposed   three-parameter models with the two-parameter  models clearly demonstrated inadequacy of the latter ones especially  when one considers approximations around the origin   in modeling daily log returns of stock market data (see also Kozubowski, 2001). 

 Improvements of these procedures using robust,  Bayesian approaches or more efficient algorithms, and the derivation of the trivariate  or joint asymptotic distribution of the three point estimators  would  be worth exploring in the future.

 \bigskip


\end{document}